\documentclass[11pt]{article}
\usepackage{hyperref}
\usepackage{amssymb}
\usepackage{graphicx}
\usepackage{slashed}
\usepackage{amsmath}
\usepackage{breqn}
\usepackage{authblk}
\usepackage{dsfont}
\usepackage{cite}
\usepackage[section]{placeins}
\usepackage[a4paper, total={7in, 9.5in}]{geometry}
\numberwithin{equation}{section}
\newcommand{\be}{\begin{equation}}
\newcommand{\ee}{\end{equation}}
\newcommand{\bea}{\begin{eqnarray}}
\newcommand{\eea}{\end{eqnarray}}
\newcommand{\ba}{\begin{aligned}}
\newcommand{\ea}{\end{aligned}}
\begin{document}
\title{ Application of regularization maps to quantum mechanical systems in 2 and 3 dimensions }
\author{E. Harikumar \thanks{eharikumar@uohyd.ac.in} and Suman Kumar Panja \thanks{sumanpanja19@gmail.com}}
\affil{School of Physics, University of Hyderabad, \\Central University P.O, Hyderabad-500046, Telangana, India}
\author{Partha Guha\thanks{ partha.guha@ku.ac.ae}}
\affil{ Department of Mathematics \\ Khalifa University of Science and Technology\\
P.O. Box 127788, Abu Dhabi, UAE}
\maketitle

\begin{abstract}
  We extend the Levi-Civita (L-C) and Kustaanheimo-Stiefel (K-S) regularization methods that maps the classical system where a particle moves under the combined influence of  $\frac{1}{r}$ and $r^2$ potentials to a  harmonic oscillator with inverted sextic potential and interactions to corresponding quantum mechanical counterparts, both in 2 and 3 dimensions. Using the perturbative solutions of the Schr\"odinger equation of the later systems, we derive the eigen spectrum of the Hydrogen atom in presence of an additional harmonic potential. We have also obtained the mapping of a particle moving in the shifted harmonic potential to H-atom using Bohlin-Sundman transformation,
for quantum regime. Exploiting this equivalence, the solution to the Schr\"odinger equation of the former is obtained from the solutions of the later.
\end{abstract}

\section{Introduction}

\smallskip

The Levi-Civita (L-C)\cite{L-C} and Kustaanheimo-Stiefel (K-S)
\cite{K-S} regularization schemes convert singular differential equations into regular ones. This is achieved by a change of variable, after re-expressing the equations in complex co-ordinates, followed by a reparametrization of time. In these schemes, existence of a constant of motion (normally, conserved energy) crucial in implementing the regularization map. Under these maps, a singular system on constant energy surface is mapped to a regular system.

These methods have been applied successfully to classical particle moving under the influence of $\frac{1}{r}$ potential in 2 and 3 dimensions, mapping these systems to harmonic oscillator in 2 and 4 dimensions, respectively\cite{L-C,K-S}. These mappings have been extended to quantum mechanical systems also, showing the equivalence between 2 and 3 dimensional H-atom and harmonic oscillator in 2 and 4 dimensions\cite{hd}. In both classical as well as quantum mechanical cases, these regularization schemes map the singular, linear differential equation to a regular linear equation. In 2 dimensions, the equations of motion of the H-atom is first expressed in terms of complex coordinates where as in 3-dimensions, it is expressed in terms of quaternions. This is then followed by a re-parametrisation of the time variable, leading to  regular, but non-linear differential equations. The non-linear terms are then re-written using the conserved energy of the former system, thereby making the equations linear. Thus the singular equations of motion describing H-atom in 2 and 3 dimensions on  constant energy surfaces are mapped to regular, linear equations describing harmonic oscillators in 2 and 4 dimensions, respectively. Thus in both classical and quantum cases, the regularization maps connect the solutions to equations of motion/Schr\"odinger equations of a fixed energy surface of the former system. Under these maps, the Hamiltonians corresponding to H-atom in 2 and 3 dimensions are mapped to harmonic oscillator Hamiltonians in 2 and 4 dimensions. These mappings between the Hamiltonians  retain this information about the constant energy surface as the strength of the harmonic potentials are  dictated by the conserved energy of the former system, both in 2 and 3 dimensions. The studies on the map between 3-dimensional H-atom and 4-dimensional harmonic oscillator brought out the role of hidden symmetries of the H-atom\cite{nmfr}.
There also exist a map between the harmonic oscillator to Kepler problem in 2-dimensions known as Bohlin-Sundman transformation \cite{bohlin}which also uses a similar procedure of first expressing the equations of motion in terms of complex coordinates, followed by a reparametrization of time variable. Here too the conserved energy of the former decides the strength of the potential of the latter system. This map has been generalized to the realm of quantum mechanics\cite{ners}.

It is of intrinsic interest to investigate the application of these regularization schemes to other quantum mechanical systems and we take up this study in this paper. It has been shown recently that the Kepler problem with an additional oscillator potential can be mapped to harmonic oscillator in presence of inverted sextic potential and interactions\cite{skp}. This map was derived in 2-dimensions using Levi-Civita(L-C) regularization \cite{L-C} and in 3-dimensions using Kustaanheimo-Stiefel(K-S) regularization \cite{K-S} schemes, respectively. The Levi-Civita map is a covering map of degree 2,  $L.C. : {\Bbb C}^2 / \{0\} \rightarrow {T^{\ast}S^2}\setminus{S^2}$. We write $(q,p)$ for coordinates of 
$T^{\ast}{\Bbb C} = {\mathds C} \times {\mathds C}$, then
in practice the Levi-Civita regularization of planar two body collisions are carried out by the variable substitution 
$(q,p) \mapsto (q^2, \frac{p}{2\bar{q}})$. Kustaanheimo-Stiefel(K.S.)  map is technically more involved. Let ${\Bbb H}$ denotes the space of quaternions. The K.S.  map is defined as
$$
K.S. : T^{\ast}({\mathds H} \setminus\{0\}) \rightarrow {\mathds H} \times {\mathds H}, \quad \hbox{s.t. } 
(z,w) \mapsto ( {\bar z}iz, \frac{{\bar z}iw}{2|z^2|}).
$$
Note that K-S regularized Hamiltonian has an additional $S^1$ symmetry resulting from a Hamiltonian $S^1$-action on
the symplectic manifold $T^{\ast}({\mathds H}\setminus \{0\})$.

Here, we generalize these two regularization schemes to the case of quantum mechanical particle moving under the influence of the combined effect of $\frac{1}{r}$ and $r^2$ potentials in 2 and 3 dimensions, which were shown to be related to harmonic oscillator with inverse sextic potential and interactions in the classical cases, recently\cite{skp}. Starting from the Schr\"odinger equation of a particle in presence of $\frac{1}{r}$ and $r^2$ potentials in 2-dimensions, we map it to the Schr\"odinger equation describing a particle in presence of harmonic, inverted sextic potentials and interactions.  The three dimensional Schr\"odinger equation describing H-atom with additional harmonic potential is shown also shown to be mapped to Schr\"odinger equation in 4-dimensions describing a particle under the influence of  harmonic, inverted sextic potentials and interactions.  In both cases, we find that the strength of sextic potential and interactions are controlled by the coefficient of the harmonic potential present in the former systems. The strength of harmonic potential in the later is governed by the conserved energy of the former systems.  We then solve these Schr\"odinger equations  by treating the sextic potential and interactions as perturbations to harmonic oscillator and derive the eigenfunctions and 
eigen values. The regularization maps, viz: Levi-Civita and K-S maps,  allows us to obtain the eigenfunctions and eigen values of the H-atom with additional harmonic potential. 

Using the regularization map between the shifted harmonic oscillator and H-atom, we obtain the mapping between the eigenfunctions and eigen values of these two systems. As in the case of usual oscillator to H-atom correspondence\cite{ners}, here too, we see that only parity even states of the shifted H-atom is mapped to that of the H-atom.

We also explicitly construct the canonical transformation that relates a generalization of Kepler Hamiltonian having a time dependence to a time independent Hamiltonian. Further, the corresponding Schr\"odinger equations are shown to be mapped to each other by a non-inertial transformation.

This paper is organized as follows. In the next section, we give a brief summary of the essential 
results concerning the mapping between 
Kepler problem augmented with a harmonic potential to harmonic oscillator with inverted sextic potential and interactions, both in 2 and 3 dimensions\cite{skp}. Here the former are singular systems where as the later are regular systems. 
We also recall the  results of generalization of he 
Bohlin-Sundman map connecting shifted harmonic oscillator to Kepler problem in 2 dimensions. In Sec.3, we derive 
the mapping between the Schr\"odinger 
equation of 2-dimensional Hydrogen atom in presence of harmonic potential and that of the harmonic 
oscillator coupled with inverted sextic 
potential and interactions in 2-dimensions. We then, by treating the inverted sextic potential 
and interactions as perturbations, 
solve the Schr\"odinger equation for harmonic oscillator and obtain eigenfunctions and eigen values. 
Using the mapping between 
the Schr\"odinger equations, we then obtain the eigenfunctions and eigen values of the Hydrogen atom augmented with a harmonic potential. We generalize this study to 3-dimensional Hydrogen 
atom  in presence of an additional oscillator potential and obtain the mapping of the corresponding Schr\"odinger equation to that of 4-dimensional harmonic oscillator coupled with inverted sextic potential and interaction in Sec.4. 
Further, we derive the eigenfunctions and eigen values from the Schr\"odinger equation of 4-dimensional harmonic oscillator with inverted sextic potential and interactions by treating these as perturbations. Using this wave function and eigen values, we derive the wave function and eigen values of Hydrogen atom in presence of harmonic potential in 3-dimensions.  In Sec.5, the mapping between Schr\"odinger equation describing shifted harmonic oscillator to H-atom is used to obtain the spectrum of the former from that of the later. Our concluding remarks are given in the last section. In the appendix A, we show that the system described in Eqn.(\ref{Xeqnn}) below is related to a generalization of the Kepler motion whose Hamiltonian has a time dependence, by a canonical transformation and then we generalize this to quantum mechanical level. In appendix B, we show that these two systems are related by a non-inertial co-ordinate transformation by mapping their Schr\"odinger equations.

\section{Review of regularization of Kepler problem in presence an additional confining potential }

In this section, we briefly recall essential results of the generalization of the Levi-Civita(L-C) regularization \cite{L-C} and Kustaanheimo-Stiefel(K-S) regularization \cite{K-S} to Kepler problem augmented with an additional oscillator potential, both  in 2-dimensions and 3-dimensions, respectively\cite{skp}. We also summarize the Bohlin-Sundman map applied to shifted harmonic oscillator in 2-dimensions. We will use these results in the next sections.

The regularization methods\cite{L-C, K-S} map linear, singular, Kepler equations in 2 and 3 dimensions to regular and linear equations describing harmonic oscillators in 2 and 4 dimensions, respectively. In \cite{skp}, it has been shown that the linear, singular equation describing the motion of a particle under the combined influence of Kepler and oscillator potential in 2 and 3 dimensions are mapped to regular, but still non-linear equations
describing harmonic oscillator with inverted sextic potential and interactions. These methods of regularization crucially depend on the existence of constant of motion(energy) and consist of re-writing equations of motion using re-parameterization of time followed by a re-definition of co-ordinates. Finally, the conserved energy expression is used to re-express certain terms appearing in the equations of motion.

We will generalize these mappings to the corresponding quantum mechanical systems in this paper.

\subsection{Regularization of motion under the combined potential of $\frac{1}{r}$ and $r^2$ in two dimensions}

In this subsection, we summarize the mapping of equations of motion of a particle moving in 
$\frac{1}{r}$ potential with an additional oscillator potential to that of a particle moving under the influence of a harmonic oscillator potential, but coupled with inverted sextic potential and interactions. 

The equation of motion of a particle moving under the combined influence of Kepler and oscillator potentials is
\bea
m \frac{d^2X_{i}}{dt^2}-\frac{m\lambda^2}{4}X_i+\frac{kX_i}{(X_{1}^2+X_{2}^2)^{\frac{3}{2}}}=0,     ~i=1,2. \label{Xeqnn}
\eea
These equations are shown to follow as Euler-Lagrange equations from the Lagrangian
\be
L=\frac{m}{2}({\dot X_1}^2+{\dot X_2}^2)+\frac{m\lambda^2}{8}(X_{1}^2+X_{2}^2)-\frac{m\lambda}{2}(X_1{\dot X_{1}}+X_2{\dot X_{2}})
+\frac{k}{r}.\label{lag1}
\ee
Here an over dot represents differentiation with respect to time $t$. Note that the third term on RHS is a total derivative term and will not contribute to equations of motion. The corresponding Hamiltonian 
\be
{\cal H}=\frac{(P_{X_{1}}^2+P_{X_{2}}^2) }{2m}+\frac{\lambda}{2} (X_1P_{X_{1}}+X_2P_{X_{2}})-\frac{k}{r}, \label{kham}
\ee
where $r=\sqrt{X_{1}^2+X_{2}^2}$ is a constant of motion. Since we consider the bounded motion, we express the negative energy
in terms of velocities and co-ordinates, viz;
\be
  -{\cal E}=\frac{m}{2}({\dot X_{1}}^2+{\dot X_{2}}^2)-\frac{\lambda^2}{8}m(X_{1}^2+X_{2}^2)-\frac{k}{r}.\label{h}
\ee
 We re-express the above conserved energy and the equation of motion in Eqn.(\ref{Xeqnn}) 
 in terms of complex variable $Z=X_1+iX_2$ and then we applied Levi-Civita regularization obtaining
 \be
  \frac{d^2U}{d\tau^2}+\frac{{\cal E}}{2mc^2}U-(\frac{3\lambda^2}{16c^2}|U|^4)U=0,
  \ee
where $Z=U^2$. Also note that we have used reparametrization of time variable and introduced a new time $\tau$ through $\frac{d}{dt}=\frac{c}{r}\frac{d}{d\tau}$, where $c$ is a proportionality constant. It is easy to see that these equations follow from the Lagrangian
  \be
  {\cal L}=\frac{m}{2}(U_{1}^{\prime 2}+U_{2}^{\prime 2})-\frac{{\cal E}}{4c^2}(U_{1}^2+U_{2}^2)
  +\frac{1}{32}\frac{m\lambda^2}{c^2}\left(U_{1}^2+U_{2}^2\right)^3,\label{secho}
  \ee
  where ${}^\prime$ denotes differentiation with respect to $\tau$. This Lagrangian describes an oscillator in 2-dimensions with inverted sextic potential and couplings. Note that, for small $\lambda$, this is
a system of uncoupled harmonic oscillators with perturbations involving couplings and inverted sextic potential. The Hamiltonian following from the above Lagrangian is given by
  \be
  H=\frac{P_{U_{1}}^2}{2m}+\frac{P_{U_{1}}^2}{2m}+\frac{{\cal E}}{4c^2}(U_{1}^2+U_{2}^2)
-\frac{1}{32}\frac{m\lambda^2}{c^2}\left(U_{1}^2+U_{2}^2\right)^3.\label{secho1}  
  \ee
Note here that the strength oscillator potential in the above is given by the conserved energy ${\cal E}$ of the former system while the strength of the inverted sextic potential and interactions are controlled by the coefficient $\lambda$ of the additional oscillator potential in the former system (see Eqn.(\ref{lag1})).  
  

\subsection{Regularization of 3-dimensional Kepler problem in presence an additional confining potential}

  Applying the Kustaanheimo-Stiefel(K-S) transformation\cite{K-S}, equivalence between the motion under the combined action of Kepler potential and an oscillator potential in three dimension and perturbed 4-dimensional Harmonic Oscillator was established\cite{skp}.

The equations of motion of a particle moving in 3-dimensions, under the action of $\frac{1}{r}$ and $r^2$ potentials is 
\be
m{\ddot X}_i-\frac{m\lambda^2}{4}X_i+\frac{k}{r^3} X_i=0,~i=0,1,2\label{eleqn}.
\ee
The Hamiltonian for this system is
\be
H=\sum_{i=0}^2\frac{P_{i}^2}{2m}+\sum_{i=0}^2\frac{\lambda}{2}X_iP_i-\frac{k}{r}\label{dkham}.
\ee
This constant of motion, in terms of velocities and positions is
\be
-{\cal E}=-\frac{E}{m}=\sum_{i=0}^2\frac{{\dot X}_{i}^2}{2}-\sum_{i=0}^2\frac{\lambda^2}{8}X_{i}^2-\frac{\mu}{r}\label{com}.
\ee
 Applying re-parameterization of time and co-ordinate transformation(K-S transformation) 
\be
\frac{d}{d t}=\frac{1}{4r}\frac{d}{d\tau} ~~ and ~~ X={\underline U}~{\underline U}^*,\label{trepara}
\ee
 where ${\underline U}$ is quaternions(see \cite{skp} for details),
the singular Eqn.(\ref{eleqn}) are mapped to
\be
{U}_{i}^{\prime\prime } +\left(8{\cal E}-3\lambda^2 |{\underline U}|^4 \right){ U}_{i}=0, ~i=0,1,2,3.\label{sho1}
\ee
Note that the $\lambda^2$ dependent term carries the signature of additional oscillator potential present in the initial system. These equations are the E-L equations following from the Lagrangian
\be
{\cal L}_{SO}=\frac{1}{2}m\sum_{i=0}^3(U_{i}^\prime)^2-4m{\cal E}\sum_{i=0}^3U_{i}^2+\frac{m\lambda^2}{2}(\sum_{i=0}^3U_{i}^2)^3\label{sextic}, ~i=0,1,2,3.
\ee
 and the corresponding Hamiltonian is 
\be
H=\sum_{i=0}^{3}\frac{\tilde{P_{i}}^2}{2m}+4m{\cal E}(\sum_{i=0}^{3}U_{i}^2)-\frac{m\lambda^2}{2}(\sum_{i=0}^{3}U_{i}^2)^3.\label{sexticH}
\ee
Note that the coefficient of the oscillator potential in the 
above i nothing but the conserved energy of the former system. 
This Lagrangian/Hamiltonian, describes a harmonic oscillator with an inverted sextic potential and interactions in 4-dimensions.

\subsection{Mapping of shifted Harmonic oscillator : Bohlin-Sundman Map}

Here we summarize the generalization of Bohlin-Sundman transformation  applied  to equations of a particle in harmonic oscillator potential whose coefficient is the shifted frequency of the oscillator. This maps the equation of motion to that of a particle moving in $\frac{1}{r}$ potential.

A time-dependent co-ordinate transformation $x_i=q_{i}e^{\frac{\lambda \tau}{2}}$ allows one to 
re-express the below equations of motion in two-dimensions,
\be
{ q}_{i}^{\prime\prime}+\lambda q_{i}^\prime+\Omega^2q_i , ~i=1,2 \label{dho-eq}
\ee
as
\be
x_{i}^{\prime\prime}+{\tilde{\Omega}}^2x_i=0,~~i=1,2\label{sho-eq}
\ee
where ${\tilde{\Omega}}^2=\Omega^2-\frac{\lambda^2}{4}$. These shifted harmonic oscillator equations follow from
the Lagrangian
\be
L=\frac{m}{2}\left( x_{1}^{\prime 2}+ x_{2}^{\prime 2}\right) -\frac{m}{2}{\tilde{\Omega}}^2(x_{1}^2+x_{2}^2)-\frac{m\lambda}{2}(x_1 x_{1}^\prime+x_2  x_{2}^\prime), \label{lag2}
\ee
where $x_{i}^{\prime}=\frac{dx_{i}}{d\tau}$ and corresponding Hamiltonian is
\be
H=\frac{1}{2m}(p_{1}^2+p_{2}^2)+   \frac{m{\Omega}^2}{2}(x_{1}^2+x_{2}^2)+\frac{\lambda}{4}(x_1p_1+p_1x_1+x_2p_2+p_2x_2). \label{ham2}
\ee
This conserved quantity in terms of velocities and coordinates becomes
\be
H=\frac{m}{2}\left( x_{1}^{\prime 2}+ x_{2}^{\prime 2}\right)+\frac{m{\tilde\Omega}^2}{2}\left(x_{1}^2+x_{2}^2\right). \label{ham3}
\ee
In terms of complex co-ordinate $\omega=x_1+ix_2$  Eqn.(\ref{sho-eq}) becomes
\be
{\omega^{\prime\prime}}+{\tilde{\Omega}}^2\omega=0.\label{comeqn}
\ee
and Eqn.(\ref{ham3}) becomes
\be
 H=\frac{m}{2}\left[ {\bar{\omega}}^\prime \omega^\prime+{\tilde\Omega}^2{\bar\omega}{\omega}\right] \equiv E. \label{ham4}
\ee
Applying the Bohlin-Sundman transformation and re-parameterization of time, i.e., 
\be
\omega\to Z=\omega^2 ~~ and ~~{\bar Z}\frac{dZ}{dt}=\frac{\bar \omega}{2}\frac{d\omega}{d\tau} \label{bohlin1}
\ee
Eqn.(\ref{comeqn}) (using Eqn.(\ref{ham4})) becomes
\be
\frac{d^2Z}{dt^2}=-\frac{E}{4m}\frac{Z}{|Z|^3}\label{eQe}.
\ee
Here $E$, the conserved energy is related to strength of Kepler potential as $k=\frac{E}{4}$.
This equation is the Kepler's equation in 2-dim, written in the complex co-ordinate $Z=X_{1}+iX_{2}$. The above equation can be shown to come from the Hamiltonian,
\be
{\cal H}=\frac{P_{X_{1}}^2}{2m}+\frac{P_{X_{2}}^2}{2m}-\frac{k}{\sqrt{X_{1}^2+X_{2}^2}}, \label{kepham}
\ee
where $k=\frac{E}{4}$.

\section{ Mapping of the Schr\"odinger Equation with $\frac{1}{r}$ and $r^2$ potentials in two dimensions}

In this section, we obtain the spectrum of the Schr\"odinger equation corresponding to the system described by the Hamiltonian in Eqn.(\ref{kham}). For this, we first map the Schr\"odinger
 equation for the Hamiltonian in Eqn.(\ref{kham}) to that of harmonic oscillator with inverted sextic potential and couplings, whose Hamiltonian is given in Eqn.(\ref{secho1}).  After solving the second of these using the perturbative methods, and using the map, we calculate the eigen values and eigenfunctions of the former system.

 For this, we start with the Hamiltonian
\be
{\cal H}=\frac{(P_{X_{1}}^2+P_{X_{2}}^2) }{2m}+\frac{\lambda}{4} (X_1P_{X_{1}}+P_{X_{1}}X_1)+\frac{\lambda}{4}(X_2P_{X_{2}} +P_{X_{2}}X_2  )-\frac{k}{r},\label{khamsy}
\ee
where we have symmetrized the $X_iP_{X_{i}}$ terms. After re-expressing above Hamiltonian as
\be
{\hat {\cal H}}=\frac{(P_{X_{1}}+\frac{m\lambda X_1}{2} )^2}{2m}+\frac{(P_{X_{2}}+\frac{m\lambda X_2}{2} )^2}{2m}-\frac{\lambda^2 m}{8}(X_{1}^2+X_{2}^2)-\frac{k}{\sqrt{X_{1}^2+X_{2}^2}} \label{khamsy1}
\ee
we set up the Schr\"odinger equation 
\be
{\hat {\cal H}}\Phi=E\Phi.
\ee
We now apply a transformation $\eta{\hat {\cal H}}\eta^{-1}\eta\Phi=E\eta\Phi$  to above equation and redefine $\eta{\hat {\cal H}}\eta^{-1}=H$ and $\eta\Phi=\psi$. Taking 
\be
\eta=exp~\left({\frac{im\lambda(X_{1}^2+X_{2}^2)}{4\hbar}}\right)
\ee
and noting $\eta(P_{X_{i}}+\frac{m\lambda X_{i}}{2})^n\eta^{-1}=P_{X_{i}}^n$, we find the transformed Hamiltonian to be
\be
H=\frac{(P_{X_{1}}^2+P_{X_{2}}^2) }{2m}-\frac{\lambda^2 m}{8}(X_{1}^2+X_{2}^2)-\frac{k}{r}.\label{hgt1}
\ee
This Hamiltonian describes a 2-dimensional H-atom with an additional inverted harmonic potential. In the polar coordinates, this Hamiltonian becomes 
\footnote{Here the Hamiltonian in Eqn.(\ref{khamsy}), in polar co-ordinate is $\hat{H}=\frac{1}{2m}(P_{r}^2+P_{\theta}^2)-\frac{\lambda}{4}(P_{r}r+rP_{r})-\frac{k}{r}$. We note that with $\eta=exp~\left({\frac{im\lambda r^{2}}{4\hbar}}\right)$, we get  $H=\eta\hat{H}\eta^{-1}$, where H is same as the one given in the Eqn.(\ref{hpolarr}). Thus, the transformation from $\hat{H}$ in Eqn.(\ref{khamsy1}) to H in Eqn.(\ref{hgt1}) can be implemented in polar co-ordinates, directly giving the Hamiltonian in Eqn.(\ref{hpolarr}).}
\be
H=\frac{1}{2m}(P_{r}^2+P_{\theta}^2)-\frac{\lambda^2}{8}mr^2-\frac{k}{r}\label{hpolarr}
\ee
and we will now find solution to the Schr\"odinger equation, in polar coordinates, i.e.,
\be
\left[ -\frac{\hbar^2}{2m}\left(\frac{\partial^2}{\partial r^2}+\frac{1}{r}\frac{\partial}{\partial r}+\frac{1}{r^2}\frac{\partial^2}{\partial\theta^2}\right)-\frac{\lambda^2 m}{8}r^2-\frac{k}{r}\right]
\psi(r,\theta)={ E}\psi(r,\theta).\label{EK-polarr}
\ee
where we have used the realization
\bea
P_r&=&-i\hbar\left(\frac{\partial}{\partial r} +\frac{N-1}{2r}\right)\\
P_\theta&=&-i\hbar\frac{\partial}{\partial \theta}.
\eea
Note that, in our case, $N=2$. in the above.

To find the solution to above equation, Eqn.(\ref{EK-polarr}), we first generalize the equivalence of above system to that described by the Schr\"odinger equation corresponding to the 
Hamiltonian in Eqn.(\ref{secho1}) and use the solution of the latter, obtained using perturbative approach. For this we re-express the Hamiltonian in Eqn.(\ref{secho1}) in polar coordinates, viz:
\be
{\tilde H}=\frac{P_{\rho}^2}{2m}+\frac{P_{\phi}^2}{2m}+\frac{m}{2}\Omega_{0}^2\rho^2-\frac{\lambda^2 m}{32c^2}\rho^6\label{sechopolar}
\ee
where we have used $\frac{m\Omega_{0}^2}{2}=\frac{\cal E}{4c^2}$. The corresponding Schr\"odinger equation is given by
\be
\left[ -\frac{\hbar^2}{2m}\left(\frac{\partial^2}{\partial\rho^2}+\frac{1}{\rho}\frac{\partial}{\partial\rho}+\frac{1}{\rho^2}\frac{\partial^2}{\partial\phi^2}\right)+\frac{m}{2}\Omega_{0}^2\rho^2-\frac{\lambda^2 m}{32c^2}\rho^6\right]
\psi(\rho,\phi)={\tilde E}\psi(\rho,\phi).\label{secos-polarr}
\ee

To see the equivalence of the Schr\"odinger equations given in Eqn.(\ref{secos-polarr}) and Eqn.(\ref{EK-polarr}), we have to apply the coordinate transformations corresponding to the Levi-Civita map ($Z=U^2$) used in 
relating these two systems, classically\footnote{One also need to implement the re-parameterization of time variables, but this will not affect the relations
$-i\hbar\frac{\partial \psi(r,\theta)}{\partial t}=E\psi(r,\theta)$ and $-i\hbar\frac{\partial \psi(\rho,\phi)}{\partial \tau}={\tilde E}\psi(\rho,\phi)$.} . That is, we apply the relations
\footnote{
Under the map, $Z=U^2$ the conserved angular momentum $J_H$ of the modified H-atom and that of perturbed oscillator  $J_{O}$ are related as $2J_H=J_O$ and this fixes $C=\frac{1}{4}$.}
\be
r=\rho^2, \theta=2\phi,~{\rm and~choose~} c=\frac{1}{4}\label{l-c-polarr}
\ee
and by a change of variables, map above Hamiltonian operators(and thus the Schr\"odinger equations), where we make further identifications
\be
E=-\frac{1}{8}m\Omega_{0}^2,~~~{\rm and~~} 4k={\tilde E}.\label{parameter}
\ee

We now solve the Schr\"odinger equation in Eqn.(\ref{secos-polarr}), by treating the $\lambda^2$ dependent term as a perturbation. After writing ${\tilde E}={\tilde E}^0+\lambda^2 g$ as the unperturbed energy, we readily 
find the energy eigen value and eigenfunction as\cite{fu}
\bea
{\tilde E}_{n_{\rho}, l}^{0}&=&\hbar\Omega_0(2n_{\rho}+|l|+1),\label{eevv1}\\
\psi_{n_{\rho}, l}^{0}(\rho,\phi)&=&\frac{1}{\sqrt{2\pi} }e^{-\frac{1}{2}\alpha\rho^2}\rho^{|l|}{}_1F_{1}(-n_{\rho}; |l|+1;\alpha\rho^2)e^{il\phi},\label{eeff}
\eea
respectively. In the above, we have defined $\alpha=m\Omega_0/\hbar$. Using the perturbative scheme, we find the first order correction to the eigen value to be
\be
{\tilde E}_{n_\rho, l}^\prime=-\frac{\lambda^2 m}{64c^2}\frac{1}{\alpha^4}
\left(\begin{array}{c}
n_\rho+|l|\\
n_\rho
\end{array}\right) \left(\begin{array}{c}
n_\rho-4\\
n_\rho
\end{array}\right)\Gamma(|l|+4){}_3F_2(-n_\rho, |l|+4, 4; |l|+1, 4-n_\rho; 1)A_{n_{\rho},l},\label{ceevv}
\ee
where $A_{n_{\rho},l}=\left(\frac{n_{\rho}!|l|!}{(n_\rho+|l|)!}\right)^2\frac{1}{\alpha^{ |l|}}$ and the first order correction to the eigenfunction is
\be
\psi_{n_{\rho},l}^\prime(\rho,\phi)=\sum_{n_{\rho}\ne n_{\rho}^\prime} \frac{A B}{({\tilde E}_{n_{\rho}}^0-{\tilde E}_{n_{\rho}^\prime})}\psi_{n_{\rho}^\prime, l^\prime}(\rho,\phi)\label{ceeff}
\ee
where 
\bea
A&=&-\frac{\lambda^2 m}{64c^2}\frac{1}{\alpha^{4+|l|}}\frac{n_{\rho}!|l|!}{(n_\rho+|l|)!}\frac{n_{\rho}^\prime !|l|!}{(n_{\rho}^\prime+|l|)!}\\
B&=&\left(\begin{array}{c}
n_{\rho}^\prime+|l|\\
n_\rho
\end{array}\right)\left(\begin{array}{c}
n_\rho-4\\
n_\rho
\end{array}\right)\Gamma(|l|+4){}_3F_{2}(-n_{\rho}^\prime, |l|+4,4;|l|+1,4-n_\rho;1).
\eea
In calculating these corrections we have used certain identities given in\cite{grad}.
Under the map, $Z=U^2$, (or equivalently, $r=\rho^2, \theta=2\phi$), conserved angular momentum $J_H$ of the modified H-atom and that of perturbed oscillator  $J_{O}$ are related as $2J_H=J_O$ and this, 
apart from fixing  $C=\frac{1}{4}$ (see Eqn.(\ref{l-c-polarr})), relates the angular momentum quantum numbers of modified H-atom, ${\tilde l}$ and that of the perturbed oscillator $l$ as 
\be
{\tilde l}=\frac{l}{2}\label{ang-momntumqn}.
\ee
This implies that $n_{r}=\frac{n_{\rho}}{2}$ i.e., when $n_{\rho}$ $\in$  $Z^{+}$ (i.e., even integer), $n_{r}$ $\in$ $Z$ and when $n_{\rho}$ $\in$ $Z^{-}$ (i.e., odd integer), 
$n_{r}$ $\in$ $\frac{Z}{2}$ $\Rightarrow$ $n_{r}$ $\sim$ half integer, which is not allowed. Thus, we see that the eigenfunctions and eigen values corresponding to even values of $n_\rho$ are mapped to the eigenfunctions and eigen values of the H-atom with additional oscillator potential. We will comment on the possible mapping of eigenfunctions and eigen values corresponding to odd integer values of $n_\rho$ in the concluding remarks.

Now, using $r=\rho^2$, $\theta=2\phi$, $l=2{\tilde l}$ and $\alpha=2\beta$ in Eqn.(\ref{eevv1},\ref{eeff},\ref{ceevv},\ref{ceeff}), we find the eigen value and eigenfunction corresponding to the Schr\"odinger equation in Eqn.(\ref{EK-polarr})
\bea
E=E_{n_{r},{\tilde l}}^0+E_{n_{r},{\tilde l}}^\prime\label{meev}\\
\psi_{n_{r},{\tilde l}}=\psi_{n_{r},{\tilde l}}^0+\psi_{n_{r},{\tilde l}}^\prime.\label{meef}
\eea
where
\bea
E_{n_{r},{\tilde l}}^0&=&-\frac{mk^2}{2\hbar^2(n_r+|{\tilde l}|+\frac{1}{2})^2}\\
E_{n_{r},{\tilde l}}^\prime&=&-\frac{\lambda^2m}{8(2\beta)^4}B_{n_{r},{\tilde l}}\left(\begin{array}{c}
n_r+2|{\tilde l}|\\
n_r
\end{array}\right)\left(\begin{array}{c}
n_r-4\\
n_r
\end{array}\right)\Gamma(2|{\tilde l}|+4)F\\
{\rm~where~} F&=&{}_3F_2(-n_r,2|{\tilde l}|+4,4;2|{\tilde l}|+1, 4-n_r;1)\\
{\rm and~}B_{n_{r}|{\tilde l}|}&=& (2\beta)^{-2|{\tilde l}|}\left(\frac{n_r!(2|{\tilde l}|)!}{(n_r+2|{\tilde l}|)!}\right)^{2}
\eea
and similarly, we find 
\bea
 \psi_{n_r,{\tilde l}}^0&=&\frac{1}{\sqrt{2\pi}}e^{i{\tilde l}\theta} r^{|{\tilde l}|}e^{-\beta r}{}_1F_1(-n_r; 2|{\tilde  l}|+1; 2\beta r)\label{eign1}\\
\psi_{n_r,{\tilde l}}^\prime&=&\sum_{n_r\ne n_{r}^\prime}\frac{{\tilde B}\left(\begin{array}{c}
n_{r}^\prime+2|{\tilde l}|\\
n_r
\end{array}
\right)\left(\begin{array}{c}
n_{r}-4\\
n_r
\end{array}
\right)\Gamma(2|{\tilde l}|+4){\tilde F}\psi_{n_{r}^\prime,{\tilde l}}^0} {E_{n_{r},{\tilde l}}^0- E_{n_{r}^\prime,{\tilde l}}^0}\label{eigg1}
\eea
where 
\bea
{\tilde B}&=&-\frac{\lambda^2m}{8(2\beta)^{4+2|{\tilde l}|}} \frac{n_r!(2|{\tilde l}|)!}{(n_r+2|{\tilde l}|)!}\frac{n_{r}^\prime !(2|{\tilde l}|)!}{(n_{r}^\prime+2|{\tilde l}|)!}\\
{\tilde F}&=& 
{}_3F_2(-n_{r}^\prime, 2|{\tilde l}|+4,4;
2|{\tilde l}|+1, 4-n_r; 1)
\eea
Note here that the Hamiltonian in Eqn.(\ref{khamsy}) and one in Eqn.(\ref{hgt1}) are gauge equivalent. Thus, measured quantities such as
 energy eigen values will be same for both Hamiltonian but energy eigenfunction will modified. From the relation $\eta\Phi=\psi$ we get energy eigenfunction of the Hamiltonian in Eqn.(\ref{khamsy}) as
\bea
\Phi = exp~\left({-\frac{im\lambda r^{2}}{4\hbar}}\right)(\psi_{n_r,{\tilde l}}^0 + \psi_{n_r,{\tilde l}}^\prime),
\eea
where $\psi_{n_r,{\tilde l}}^0$ and $\psi_{n_r,{\tilde l}}^\prime$ are given in Eqn.(\ref{eign1}) and Eqn.(\ref{eigg1}), respectively.


\section{ Mapping of the Schr\"odinger Equation with potentials 
$\frac{1}{r}$ and $r^2$  in three dimensions }

Here, in this section, we obtain the eigen values and eigenfunctions of 
the Schr\"odinger equation corresponding to the Hamiltonian in Eqn.(\ref{dkham}). This is derived from the perturbative solution of Schr\"odinger 
equation corresponding to Eqn.(\ref{sexticH}), by first establishing the equivalence between these two Schr\"odinger equations.

For this, we start with the Hamiltonian
\be
{\cal H}=\frac{(P_{X_{0}}^2+P_{X_{1}}^2+P_{X_{2}}^2) }{2m}+\frac{\lambda}{4} (X_0P_{X_{0}}+P_{X_{0}}X_0)+\frac{\lambda}{4} (X_1P_{X_{1}}+P_{X_{1}}X_1)+\frac{\lambda}{4}(X_2P_{X_{2}} +P_{X_{2}}X_2  )-\frac{k}{r},\label{khamsym}
\ee
where we have symmetrized the $X_iP_{X_{i}}$ terms. After re-expressing above Hamiltonian as
\be
{\hat {\cal H}}=\frac{(P_{X_{0}}+\frac{m\lambda X_0}{2} )^2}{2m}+\frac{(P_{X_{1}}+\frac{m\lambda X_1}{2} )^2}{2m}+\frac{(P_{X_{2}}+\frac{m\lambda X_2}{2} )^2}{2m}-\frac{\lambda^2 m}{8}(X_{0}^2+X_{1}^2+X_{2}^2)-\frac{k}{\sqrt{X_{0}^2+X_{1}^2+X_{2}^2}} \label{khamsym1}
\ee
we set up the Schr\"odinger equation 
\be
{\hat {\cal H}}\Phi=E\Phi.
\ee
We now apply a transformation $\eta{\hat {\cal H}}\eta^{-1}\eta\Phi=E\eta\Phi$  to above equation and redefine $\eta{\hat {\cal H}}\eta^{-1}=H$ and $\eta\Phi=\psi$. Taking 
\be
\eta=exp~\left({\frac{im\lambda(X_{0}^2+X_{1}^2+X_{2}^2)}{4\hbar}}\right)
\ee
and noting that $\eta(P_{X_{i}}+\frac{m\lambda X_{i}}{2})^n\eta^{-1}=P_{X_{i}}^n$, we find the transformed Hamiltonian to be
\be
H=\frac{(P_{X_{0}}^2+P_{X_{1}}^2+P_{X_{2}}^2) }{2m}-\frac{\lambda^2 m}{8}(X_{0}^2+X_{1}^2+X_{2}^2)-\frac{k}{r}.\label{hgt}
\ee
This Hamiltonian describes a 3-dimensional H-atom with an additional inverted harmonic potential. In the spherical polar coordinates, this Hamiltonian becomes 
\footnote{Here the Hamiltonian in Eqn.(\ref{khamsym}), in spherical-polar co-ordinate is $\hat{H}=\frac{1}{2m}(P_{r}^2+P_{\theta}^2+P_{\phi}^2)-\frac{\lambda}{4}(P_{r}r+rP_{r})-\frac{k}{r}$. We note that with $\eta=exp~\left({\frac{im\lambda r^{2}}{4\hbar}}\right)$, we get  $H=\eta\hat{H}\eta^{-1}$, where H is same as the one given in the Eqn.(\ref{hpolar}). Thus, the transformation from $\hat{H}$ in Eqn.(\ref{khamsym1}) to H in Eqn.(\ref{hgt}) can be implemented in spherical-polar co-ordinates, directly giving the Hamiltonian in Eqn.(\ref{hpolar}).}
\be
H=\frac{1}{2m}(P_{r}^2+P_{\theta}^2+P_{\phi}^2)-\frac{\lambda^2}{8}mr^2-\frac{k}{r}\label{hpolar}
\ee
and we will now find solution to the Schr\"odinger equation, in spherical-polar coordinates, i.e.,
\be
\left[ -\frac{\hbar^2}{2m}\left(\nabla_{3}^{2}\right)-\frac{\lambda^2 m}{8}r^2-\frac{k}{r}\right]
\Psi(r,\theta)={ E}\Psi(r,\theta).\label{EK-polar}
\ee
where $\nabla_{3}^{2}$ is Laplacian in 3-dimensional spherical coordinates, i.e,

$\nabla_{3}^{2}=\frac{1}{r^2}\frac{\partial}{\partial r}(r^2 \frac{\partial}{\partial r})+\frac{1}{r^2sin\theta}\frac{\partial}{\partial \theta}(sin \theta \frac{\partial}{\partial \theta})+\frac{1}{r^2 sin^{2}{\theta}}\frac{\partial^{2}}{\partial \phi^{2}}$

To find the solution to above equation, Eqn.(\ref{EK-polar}), we first generalize the equivalence of above system to that described by the Schr\"odinger equation corresponding to the 
Hamiltonian in Eqn.(\ref{sexticH}) and use the solution of the latter, obtained using perturbative approach. For this we re-express the Hamiltonian in Eqn.(\ref{sexticH}) in polar coordinates, viz:
\be
{\tilde H}=\frac{P_{\rho}^2}{2m}+\frac{P_{\theta}^2}{2m}+\frac{P_{\psi}^2}{2m}+\frac{P_{\phi}^2}{2m}+\frac{m}{2}\omega_{0}^2\rho^2-\frac{\lambda^2 m}{2}\rho^6 , \label{sechopolar}
\ee
where we have used $\frac{\omega_{0}^2}{2}=4{\cal E}$. The corresponding Schr\"odinger equation is given by
\be
\left[ -\frac{\hbar^2}{2m}\left(\nabla_{4}^{2}\right)+\frac{m}{2}\omega_{0}^2\rho^2-\frac{\lambda^2 m}{2}\rho^6\right]
\Psi(\rho,\theta,\psi,\phi)={\tilde E}\Psi(\rho,\theta,\psi,\phi),\label{secos-polar}
\ee
where $\nabla_{4}^{2}$ is Laplacian in 4-dimensional hyper-spherical coordinates (see this in \cite{arda}), i.e,

$\nabla_{4}^{2}=\frac{1}{r^3}\frac{\partial}{\partial r}(r^3 \frac{\partial}{\partial r})+\frac{1}{r^2sin^{2}\psi}\frac{\partial}{\partial \psi}(sin^{2} \psi \frac{\partial}{\partial \psi})+\frac{1}{r^2 sin^{2}{\psi}sin\theta}\frac{\partial}{\partial \theta}(sin \theta \frac{\partial}{\partial \theta})+\frac{1}{r^2 sin^{2}{\psi}sin^{2}\theta}\frac{\partial^{2}}{\partial \phi^{2}}$

To see the equivalence of the Schr\"odinger equations given in Eqn.(\ref{secos-polar}) and Eqn.(\ref{EK-polar}), we use K-S transformation of co-ordinates in Eqn.(\ref{trepara})  in 
relating these two systems\footnote{One also need to implement the re-parameterization of time variables, but this will not affect the relations
$-i\hbar\frac{\partial \psi(r,\theta,\phi)}{\partial t}=E\psi(r,\theta,\phi)$ and $-i\hbar\frac{\partial \psi(\rho,\theta,\psi,\phi)}{\partial \tau}={\tilde E}\psi(\rho,\theta,\psi,\phi)$.} . That is, we apply the relations

\be
r=\rho^2,~ \nabla_{3}^{2}\Psi=\frac{1}{4r}\nabla_{4}^{2}\Psi \label{l-c-polar}
\ee
and by a change of variables, map above Hamiltonian operators(and thus the Schr\"odinger equations), where we make further identifications
\be
E=-\frac{1}{8}m\omega_{0}^2,~~~{\rm and~~} 4k={\tilde E} \label{parameter}
\ee
and a constraint $\hat{O}\Psi=0$, where $\hat{O}=U_{3}\tilde{\partial_{0}}-U_{2}\tilde{\partial_{1}}+U_{1}\tilde{\partial_{2}}-U_{0}\tilde{\partial_{3}}$,

We now solve the Schr\"odinger equation in Eqn.(\ref{secos-polar}) perturbatively. First we find energy eigen value and eigenfunction for 4-dimensional harmonic oscillator as (see \cite{kibler}) 
\bea
{\tilde E}_{N,K}^{0}&=& (N+2K+1)\hbar \omega_{0},  \label{eev1}\\
\Psi_{NLMK}(\rho,\theta,\psi,\phi)&=& A_{NLMK}\rho^{N-1}Y_{NLM}(\theta,\psi,\phi)e^{-\frac{\alpha}{2}\rho^{2}}L_{N+K}^{N}(\alpha \rho^{2})\label{eef}
\eea
respectively. Here,
\be 
 A_{NLMK}=2^{\frac{1}{2}}\alpha^{\frac{N+1}{2}}(K!)^{\frac{1}{2}}[(N+K)!]^{-\frac{3}{2}},
\ee 
 where $N \in N-{0}$, $K \in N$, $L=0,1,....N-1,$ and $M=-L,-L+1,....,L$.

In the above, we have defined $\alpha=m\omega_{0}/\hbar$. Using the perturbative method, we find the first order correction to the eigen value to be
\be
{\tilde E}_{N,K}^\prime= C
\left(\begin{array}{c}
2N+K\\
N+K
\end{array}\right) \left(\begin{array}{c}
N+K-4\\
N+K
\end{array}\right)\Gamma(N+4){}_3F_2(-(N+K), N+4, 4; N+1, 4-N-K; 1),\label{ceev}
\ee
where $C=-\frac{\lambda^2 m}{4}\frac{1}{\alpha^{(4+N)}}A_{NLMK}^2 $
and the first order correction to the eigen function is
\be
\Psi_{NLMK}^{\prime}=\sum_{K \ne K^{\prime}} \frac{AB}{(\tilde{E}_{N,K}^0-\tilde{E}_{N,K^{\prime}}^0)}\Psi_{NLMK^{\prime}}\label{ceef}
\ee
where 
\bea
A&=&-\frac{\lambda^2 m}{4}\frac{1}{\alpha^{4+N}}A_{NLMK^{\prime}}A_{NLMK}\left(\begin{array}{c}
2N+K^{\prime}\\
N+K^{\prime}
\end{array}\right)\left(\begin{array}{c}
N+K-4\\
N+K
\end{array}\right)\Gamma(N+4) \\
B&=& {}_3F_{2}(-(N+K^{\prime}), N+4,4;N+1,4-(N+K);1).
\eea
In calculating these corrections we have used certain identities given in\cite{grad}.

Now, applying K-S transformation in Eqn.(\ref{eev1},\ref{eef},\ref{ceev},\ref{ceef}), we obtain the  eigenfunction for the Schr\"odinger equation in Eqn.(\ref{EK-polar})

\be
\Psi_{nlm}(Hydrogen)=\sum_{N,L,M,K}I(nlmNLMK)(\Psi_{NLMK}+\Psi_{NLMK}^{\prime}),
\ee
where wave function for perturbed hydrogen atom problem in 3-dim, $\Psi_{nlm}=\Psi_{nlm}^{0}+\Psi_{nlm}^{\prime}$ can be calculated using above equation by following the method discussed in \cite{kibler}. Also using Eqn.(\ref{parameter}), Eqn.(\ref{eev1}) and Eqn.(\ref{ceev}) we obtained eigenvalues for perturbed hydrogen-atom problem (see details \cite{kibler},\cite{bergman}),
\be
E_{n}(Hydrogen)=-\frac{k^2m}{2{\hbar}^2n^2}+D^2\tilde{E}_{N,K}^{\prime}, ~~n=1,2...
\ee
with identification $N+2K+1=2n$ and here $D=\sum_{N,L,M,K}I(nlmNLMK)$, which can be explicitly calculated by using method discussed in \cite{kibler}.

Here the Hamiltonian in Eqn.\ref{khamsym} and in Eqn.\ref{hgt} are gauge equivalent. Thus, all measured quantities like energy eigen values will remain same for both Hamiltonian. From the relation $\eta\Phi=\Psi$ we get energy eigenfunction of the Hamiltonian in Eqn.(\ref{khamsym}) as
\bea
\Phi = exp~\left({-\frac{im\lambda r^{2}}{4\hbar}}\right)(\Psi_{nlm}^0 + \Psi_{nlm}^\prime).
\eea

\section{ Mapping of the Schr\"odinger Equation of shifted harmonic oscillator}
In the subsection.(2.3), we have summarized the study of mapping the equations of motion of a shifted harmonic oscillator to that of Kepler problem via the equations of motion of a shifted harmonic oscillator. Now, we will
derive the relation between the Schr\"odinger equations corresponding to these systems. Mapping of Schr\"odinger equations harmonic oscillator and H-atom has been derived in \cite{ners} 
and we adapt this results for the shifted oscillator we have studied in \cite{skp}.

The Hamiltonian corresponding to the shifted oscillator(\ref{ham2}) is 
\be
H=\frac{1}{2m}(p_{1}^2+p_{2}^2)+   \frac{m{\Omega}^2}{2}(x_{1}^2+x_{2}^2)+\frac{\lambda}{4}(x_1p_1+p_1x_1+x_2p_2+p_2x_2).\nonumber
\ee
As earlier, we transform this Hamiltonian to an equivalent one, by the transformation ${\hat H}=\eta H\eta^{-1}$ where $\eta=e^{\frac{i}{4\hbar}m\lambda(x_{1}^2+x_{2}^2)}$ and obtain
\be
{\hat H}= \frac{1}{2m}(p_{1}^2+p_{2}^2)+\frac{m{\tilde\Omega}}{2}(x_{1}^2+x_{2}^2),
\ee
which is the Hamiltonian for harmonic oscillator with frequency ${\tilde\Omega}=(\Omega^2-\frac{\lambda^2}{4})$. The eigenfunction satisfying  
the corresponding Schr\"odinger equation, in polar co-ordinates 
\be
\left[-\frac{\hbar^2}{2m}\left(\frac{\partial^2}{\partial\rho^2}
+\frac{1}{\rho}\frac{\partial}{\partial\rho}+\frac{1}{\rho^2}\frac{\partial^2}{\partial\phi^2}\right)
+\frac{m}{2}{\tilde\Omega}^2\rho^2\right]\psi=E\psi\label{eveqn}
\ee
is
\be
\psi=\frac{1}{\sqrt{2\pi}}e^{il\phi}\psi_{lE}(\rho)
\ee
where 
\be
\psi_{lE}(\rho)=e^{-\frac{1}{2}\alpha\rho^2}\rho^{|l|}F(-n_\rho, |l|+1, \alpha\rho^2).\label{wf}
\ee
Here $F(a,b,c)$ is Confluent Hypergeometric Function and we have used $\alpha=m{\tilde \Omega}/\hbar$. The eigen value is given by
\be
E=\hbar{\tilde\Omega}(2n_\rho+|l|+1)\label{eev}
\ee

Next, we apply the Bohlin-Sundman map (see Eqn.(\ref{bohlin1})) to above equation, i,e., in polar co-ordinate, we apply
\be
\rho^2\to r ;
\phi\to \frac{\theta}{2} ;
\alpha\to 2\beta \label{BSpolar}
\ee
and after simple algebra, get the eigen value equation to be
\be
\left[-\frac{\hbar^2}{2m}\left(\frac{\partial^2}{\partial r^2}+\frac{1}{r}\frac{\partial}{\partial r}+\frac{1}{r^2}\frac{\partial^2}{\partial\theta^2}\right)+\frac{m{\tilde\Omega}^2}{8}\right]\psi(r,\theta)=
\frac{E}{4r} \psi(r,\theta)
\ee
Substituting $\psi(r,\theta)=\frac{1}{\sqrt{2\pi}}e^{i{\tilde l}\theta}\psi(r)$, we re-write the above equation as
\be
\left[-\frac{\hbar^2}{2m}\left(\frac{\partial^2}{\partial r^2}+\frac{1}{r}\frac{\partial}{\partial r}\right)-\frac{{\tilde l}^2\hbar^2}{2mr^2}-\frac{E}{4r}\right]\psi(r)
=-\frac{m{\tilde\Omega}^2}{8}\psi(r)
\ee
Now using Eqn.(\ref{bohlin1}) and renaming $E_k$ as $E_H$ with the identification $E/4=k$, we re-express above equation as
\be
\left[-\frac{\hbar^2}{2m}\left(\frac{\partial^2}{\partial r^2}+\frac{1}{r}\frac{\partial}{\partial r}\right)-\frac{{\tilde l}^2\hbar^2}{2mr^2}-\frac{k}{r}\right]\psi(r)
=E_{H}\psi(r).\label{wwf1}
\ee
We now easily identify this as the Schr\"odinger equation describing H-atom. Since we have 
$2\phi=\theta$ (see Eqn.(\ref{BSpolar})), we find that the orbital quantum numbers are 
related as $l=2{\tilde l}$. Using this and Eqn.(\ref{BSpolar}) in Eqn.(\ref{wf}) we find the eigenfunction 
$\psi(r)$ of Eqn.(\ref{wwf1}) and thus obtain
\be
\psi(r,\theta)=\frac{1}{\sqrt{2\pi}}e^{i{\tilde l}\theta} e^{-\beta r}r^{|{\tilde l}|}F(-n_r, 2|{\tilde l}|+1, 2\beta r)\label{h-wf}
\ee
where $\beta=m{\tilde\Omega}/{2\hbar}$. From Eqn.(\ref{eev}), we also find the eigen value to be
\be
E_H=-\frac{mk^2}{2\hbar^2}\frac{1}{(n_r+|{\tilde l}|+\frac{1}{2})^2}
\ee
Since ${\tilde l}=\frac{l}{2}$, we note that of the eigenfunctions in Eqn.(\ref{wf}) only those 
corresponding to even  values of $n_\rho$ are mapped to the eigenfunctions $\psi(r,\theta)$ given in Eqn.(\ref{h-wf}), as in\cite{ners}. Note here that states with even parity of shifted harmonic oscillator get to mapped to wave function of Hydrogen atom.

\section{Conclusion}

We have showed the equivalence between the Schr\"odinger equation corresponding to $\frac{1}{r}$ 
potential augmented with a harmonic oscillator potential in 2 and 3 dimensions to the Schr\"odinger equation corresponding to a particle moving under the influence 
of harmonic potential with an additional inverted, sextic potential and interactions in 2 and 4 dimensions, respectively. 
In both cases, the coefficient of the additional oscillator potential controls the strength of inverted sextic potential and interactions. For both case, using perturbative solution of the later, we obtain the eigenfunctions and eigen values of the former.  Our results reduce to the standard 
results of the mapping between H-atom and harmonic oscillator in the vanishing limit of 
the co-efficient of additional oscillator potential, i.e., $\lambda\to 0$.

We then extend the mapping of motion of shifted harmonic oscillator to Kepler problem to the mapping between the corresponding 
Schr\"odinger equations. Using this, we map eigenfunctions and eigen values of these two systems. This map shows that the eigenfunctions of (shifted) harmonic oscillator
corresponding to even eigen values are mapped to that of H-atom, as in\cite{ners}. It has been shown in \cite{ners} that the eigenfunctions corresponding to
the odd integer eigen values are mapped to that of a charged vortex system. Here too, we note that the eigen value equation in Eqn.(\ref{eveqn}) in terms of the complex co-ordinate 
$\omega=x_1+ix_2=\rho e^{i\phi}$ is 
$(4\frac{\partial^2}{\partial_\omega \partial_{\bar\omega}}+\frac{m{\tilde\Omega}}{2}{\bar\omega}\omega)
\psi(\omega, {\bar\omega})=E\psi(\omega, {\bar\omega})$. Under the parity transformation, $\omega\to\pm\omega$ and 
we have $z\to z$. Thus we see that under  $\omega\to\pm\omega$, $arg(z)$ changes by $2\pi$. Thus parity eigenstates
$\psi_\sigma(\omega, {\bar\omega})=\psi_\sigma(\omega, {\bar\omega})e^{2i\sigma arg(\omega)}$, where $\sigma=0$ for 
even and $\sigma=\frac{1}{2}$ for odd states, respectively, are mapped to 
$\psi_\sigma({\bar z}, z)e^{i\sigma arg(z)}$. The parity even states, having zero phase factor will get mapped to 
the eigenfunctions of H-atom. The non-zero phase factor for the parity odd states will lead 
to the mapping of these states to that of charged magnetic vortex system as in\cite{ners}. This discussion is relevant for the mapping of eigenfunctions and eigen values obtained in section 3 and 4. Here too, we have seen that only the eigenfunctions corresponding to the even eigen values of harmonic oscillator with inverted sextic potential are mapped to that 
of the system with combination of $\frac{1}{r}$ potential and harmonic potential. It is thus natural to expect that the eigenfunctions corresponding to odd integer eigen values will be related to a H-atom with an additional oscillator potential in presence of a magnetic vortex.

Note that both the L-C and K-S regularization schemes map the solution of Kepler problem on a constant energy surface to that of 
harmonic oscillator. In the systems studied here, we have augmented the $\frac{1}{r}$ potential with an additional $r^2$ potential and here also this constant energy surface plays the same role. We have seen that the regular system we obtained in 2 and 4 dimensions had the coefficient of harmonic oscillator potential given by the conserved energy of the initial systems. This fact was crucial in connecting the eigen values of the corresponding Schr\"odinger equations. For the classical Kepler problem, this restriction of constant energy surface was relaxed in the Moser-regularization\cite{moser} and the Ligon-Schaaf regularization\cite{ls}. It will be of interest to see their generalization to the quantum systems studied here.

 In the appendices we show the connection between the damped systems to the system described by Eqn.(\ref{Xeqnn}) using canonical transformation as well as using a non-inertial transformation of co-ordinates.

{\bf Acknowledgments}

SKP thank UGC, India for support through JRF scheme(id.191620059604). 
Work by the author PG was supported by the Khalifa University of Science and
Technology under grant number FSU-2021-014.

\renewcommand{\thesection}{Appendix : A}
\section{}
\renewcommand{\thesection}{A}
In this appendix, we show that the system described by the Eqn.(\ref{Xeqnn}) is related by a canonical transformation to a 
dammed system\cite{damp}.


The Hamiltonian given in Eqn.({\ref{kham})
\be
{\cal H}=\frac{(P_{X_{1}}^2+P_{X_{2}}^2) }{2m}+\frac{\lambda}{2} (X_1P_{X_{1}}+X_2P_{X_{2}})-\frac{k}{\sqrt{X_{1}^{2}+X_{2}^2}}\nonumber
\ee 
is mapped to the Hamiltonian\footnote{which is derived from the Bateman-Caldirola-Kanai\cite{bck} type Lagrangian given by 
$L=e^{\lambda t}\left [\frac{m}{2}( {\dot x_{1}}^2+{\dot x_{2}}^2)+\frac{ke^{-\frac{3\lambda t}{2}}} {r}\right]$ where $r=\sqrt{x_{1}^2+x_{2}^2},$ $m=\frac{m_1+m_2}{m_1 m_2}$ and $k=Gm_1m_2$}
\be
H=\frac{e^{-\lambda t}(p_{x_{1}}^2+p_{x_{2}}^2) }{2m}-\frac{k e^{\frac{-\lambda t}{2}}}{\sqrt{x_{1}^{2}+x_{2}^{2}}},\label{h2}
\ee 
by the canonical transformation generated by $F_{2}(x_{i},P_{X_{i}})= x_{i}e^{\frac{\lambda t}{2}}P_{X_{i}}$, as ${\cal H}(X_{i},P_{X_{i}}) = H(x_{i},p_{x_{i}}) + \frac{\partial F_{2}}{\partial t}$. Here we have used the relations obtained from this generating function, 
\be
 p_{x_{i}}=e^{\frac{\lambda t}{2}}P_{X_{i}},~~ X_{i}=x_{i}e^{\frac{\lambda t}{2}}.
 \ee
The corresponding Hamiltonian operators are related by\cite{Kim}
\be
{\cal H}(\hat{X_{i}},\hat{P_{X_{i}}}) = H(\hat{x_{i}},\hat{p_{x_{i}}}) + \frac{\partial F_{2}(\hat{x_{i}},\hat{P_{X_{i}}})}{\partial t},
\ee
where we use $\hat{}$ indicate the operator nature explicitly. This can be expressed using an unitary operator as (see\cite{Kim})
\be
{\cal H}(\hat{X_{i}},\hat{P_{X_{i}}}) = H(\hat{x_{i}},\hat{p_{x_{i}}}) + i\hbar \hat{U}\frac{\partial \hat{U}^{\dagger}}{\partial t},
\ee
where $\hat{U}= e^{\frac{i}{2\hbar}\left[(\hat{x_{i}}\hat{P_{X_{i}}}+\hat{x_{i}}\hat{P_{X_{i}}})e^{\frac{\lambda t}{2}}\right]}$.
\renewcommand{\thesection}{Appendix : B}
\section{}
\renewcommand{\thesection}{B}

In this appendix, we show that a non-inertial transformation relates the model given in Eqn.(\ref{Xeqnn}) to a damped system.

The time dependent Schr\"odinger equation corresponding to the Hamiltonian 
 in Eqn.(\ref{kham}) is
\be
\left[-\frac{\hbar^{2}}{2m}(\frac{\partial^{2}}{\partial X_{i}^{2}}) -i\hbar\frac{\lambda}{4}(X_{i}\frac{\partial}{\partial X_{i}}+(\frac{\partial}{\partial X_{i}})X_{i}) -\frac{k}{\sqrt{X_{1}^2+X_{2}^2}}\right]\psi(X_{i},t)= i\hbar \frac{\partial \psi(X_{i},t)}{\partial t}
\ee\label{newham}
Now we will apply non-inertial transformation 
\be
X_{i}\to x_i=X_{i}e^{-\frac{\lambda t}{2}},~~ t=\tilde{t}
\ee
the above Schr\"odinger equation, mapping it to
\be
\left[-\frac{\hbar^{2}}{2m}(e^{-\lambda \tilde{t}}\frac{\partial^{2}}{\partial x_{i}^{2}})-\frac{k e^{- \lambda \tilde{t}}}{\sqrt{x_{1}^2+x_{2}^2}}\right]{\tilde\psi}(x_{i}, \tilde{t})= 
i\hbar \frac{\partial{\tilde\psi}(x_{i}, \tilde{t})}{\partial \tilde{t}} 
\ee
where ${\tilde\psi}(x_{i}, \tilde{t})=\psi(x_{i}e^{-\lambda \tilde{t}},\tilde{t})$. In obtaining above mapping, we have used\cite{t1}
\bea
\frac{\partial}{\partial t}= \frac{\partial}{\partial \tilde{t}} - \frac{\lambda}{2}x_i\frac{\partial}{\partial x_{i}} ; \hspace{.2cm}   \frac{\partial}{\partial X_{i}}= e^{-\frac{\lambda \tilde{t}}{2}}\frac{\partial}{\partial x_{i}}.
\eea
Note that this is the Schr\"odinger corresponding to the Hamiltonian given in Eqn.(\ref{h2}), showing the mapping between the quantum systems under the above non-inertial transformation.

\end{document}